\documentclass{PoS}
\usepackage{amsmath,amssymb,graphicx,epsfig}

\title{Spectroscopy of light tetraquark states}

\ShortTitle{Spectroscopy of light tetraquark states}

\author{\speaker{Sasa Prelovsek}\\
        Jozef Stefan Institute and University of Ljubljana, 1000 Ljubljana, Slovenia\\
        E-mail: \email{sasa.prelovsek@ijs.si}}

\author{Christian B. Lang, Markus Limmer  and Daniel Mohler \\
        Institut f\"{u}r Physik, FB Theoretische Physik, Universit\"{a}t Graz, A-8010 Graz, Austria\\
}
\author{Terrence Draper and Keh-Fei Liu\\
     Department of Physics and Astronomy, University of Kentucky, Lexington, KY 40506, USA
}
\author{Nilmani Mathur\\
     Department of Theoretical Physics, Tata Institute of Fundamental Research, Mumbai, India
}

\abstract{We address the question whether the lightest scalar mesons $\sigma$ and  $\kappa$ are  tetraquarks, as is strongly supported by many phenomenological studies.  We present a search for possible light tetraquark states with $J^{PC}=0^{++}$ and $I=0,~1/2,~3/2,~2$ on the lattice. The spectrum is determined using the generalized eigenvalue method with a number of tetraquark interpolators at the source and the sink. In all the channels, we unavoidably find lowest  scattering states $\pi(k)\pi(-k)$ or $K(k)\pi(-k)$ with back-to-back momentum $k=0,~\tfrac{2\pi}{L}~,..~$. However, we find an additional light state in the $I=0$ and $I=1/2$ channels, which may be related to the observed resonances $\sigma$ and $\kappa$ with a strong tetraquark component.  In the exotic repulsive channels $I=2$ and $I=3/2$, where no resonance is observed, we find no light state in addition to the scattering  states. 
}

\FullConference{The XXVII International Symposium on Lattice Field Theory\\
		 July 26-31, 2009\\
		 Peking University, Beijing, China}

\begin{document}

\section{Introduction}

So far, the only well established hadron states are mesons ($\bar qq$) and baryons ($qqq$). In spite of all the efforts, no exotic states like tetraquark ($\bar q\bar q qq$), pentaquark ($\bar qqqqq$), or hybrid ($\bar qqG$) states have  been confirmed beyond doubt.  Perhaps the most prominent tetraquark candidate is the $Z^+(4430)$ resonance, discovered by Belle \cite{Zbelle}: it decays to $\pi^+\psi^\prime$, so it must have a minimal quark content $\bar du\bar c c$, but it has not been confirmed by Babar (yet) \cite{Zbabar}. 

Here we address the question whether the lightest scalar mesons $\sigma$ ($I=0$), $\kappa$ ($I=1/2$) and $a_0(980)$ ($I=1$)  are tetraquarks\footnote{By tetraquark, we have in mind a state that has a dominant tetraquark Fock component.}, as many scientists strongly believe  since  Jaffe's proposal in 1977 \cite{jaffe} (see for example [3]). The $\sigma$ resonance ($m_\sigma\simeq 400-600$ MeV) is now widely accepted since its pole was determined  in a model-independent way \cite{leutwyler}.  The pole for $\kappa$ resonance ($m_\kappa\simeq 600-800$ MeV) was determined in a similar manner  \cite{descotes}. Both resonances  have been recently experimentally confirmed  \cite{scalar}, but they remain slightly controversial.    

The observed ordering $m_\kappa<m_{a_0(980)}$ can not be reconciled with the conventional $\bar us$ and $\bar ud$ states since  $m_{\bar us}>m_{\bar ud}$ is expected due to $m_s>m_d$. In the tetraquark case, the $I=1$ state $[\bar d\bar s][us]$ with additional valence pair $\bar ss$ is naturally heavier than the $I=1/2$ state $[\bar s\bar d][du]$   and the resemblance with the observed ordering favors the  tetraquark interpretation. 

Light scalar tetraquarks have been extensively studied in phenomenological models \cite{maiani, jaffe}, but there have been only few lattice simulations \cite{liu,prelovsek,tetra_lat}. The main obstacle for identifying  possible tetraquarks  on the lattice is the presence of the scattering  contributions in the correlators. The strongest claim for $\sigma$ as tetraquark was obtained for $m_\pi\simeq 180-300$ MeV by analyzing a single correlator using the sequential empirical Bayes method \cite{liu}. This result needs confirmation using a different method (for example  the variational method used here) before one can claim the existence of light tetraquarks on the lattice with some confidence.  

We determine a spectrum of states with $J^{PC}=0^{++}$, $\vec p=\vec 0$ and $I=0,~2,~1/2,~3/2$ on the lattice using the variational method with a number of  tetraquark sources and sinks. The two-pseudoscalar scattering states $\pi(k)\pi(-k)$, which have back-to-back three-momentum $k=N\tfrac{2\pi}{L}$, unavoidably couple to our sources with $I=0,~2$ and appear in the resulting spectrum. Similarly, $K(k)\pi(-k)$ appear in the $I=1/2,~3/2$ spectrums. We don't consider the more challenging $I=1$ channel, since there are two towers $\bar K(k)K(-k)$ and $\pi(k)\eta(-k)$ of scattering states. In the non-interacting limit, the scattering states have energies $ \sqrt{m_{\pi}^2+\vec k^2}+\sqrt{m_{\pi,K}^2+\vec k^2}~$. 
Our main question is whether there is any  light state with $I=0$  in addition to  the $\pi(k)\pi(-k)~$ tower. We indeed find an indication for an additional light state in $I=0$  channel, which could possibly be interpreted as $\sigma$  with strong tetraquark component. 
We also find an additional light state in $I=1/2$ channel on top of the $K(k)\pi(-k)$ tower, and this  state could be interpreted as $\kappa$ with strong tetraquark component. 
We also study the exotic repulsive channels with $I=2$ and $3/2$, where no resonance is observed.  Indeed, in this case we only find the $\pi(k)\pi(-k)$  and $K(k)\pi(-k)$ towers, but no additional light states. 

\newpage

\section{Lattice simulation}

We present results from two simulations: 
\begin{itemize}
\item An $N_f=2$ dynamical simulation with Chirally Improved quarks,  three pion masses $m_\pi=$318, 469, 526 MeV (\footnote{We analyzed $200$ configurations for lighter two pion masses and $100$ configurations for the  heaviest pion mass.}), $a\simeq 0.15$ fm and  $V=16^3\times 32$ \cite{ci}. ``Narrow'' Jacobi smearing is applied to all quark sources and sinks \cite{ci}.
\item A quenched simulation with valence overlap fermions,  $m_\pi\simeq 230-630$ MeV, $a\simeq 0.200(3)$ fm and $V=16^3\times 28$ \cite{kentucky}. In this case, quarks have point-sources and point-sinks and we analyze 300 configurations.  
\end{itemize}

We use $r=5$ tetraquark interpolators in $I=0$ and $I=1/2$ channels:
\begin{equation}
\label{basis_zero}
{\cal O}_1=PP^\prime~,\quad {\cal O}_2=\!\!\! \sum_{i=1,2,3}\!\!\! V_iV^\prime_i~,\quad {\cal O}_3=\!\!\!\sum_{i=1,2,3}\!\!\! A_iA^\prime_i~,\quad {\cal O}_4=[\bar q_1C\gamma_5 \bar q_2][q_3C\gamma_5 q_4]~,\quad {\cal O}_4=[\bar q_1C \bar q_2][q_3C q_4]~,
\end{equation}
where $[q_3C\gamma_5 q_4]$ and $[q_3C q_4]$ are (pseudo)scalar di-quarks and $P\equiv \bar q_1\gamma_5 q_2$, $V_i\equiv \bar q_1\gamma_i q_2$, $A_i\equiv \bar q_1\gamma_i \gamma_5 q_2$ are conventional currents. Appropriate quark flavor combinations\footnote{The current-current interpolators have flavor structure $2\bar du\bar u d-\bar u u\bar dd+\tfrac{1}{2}\bar uu\bar uu+\tfrac{1}{2}\bar dd\bar dd$ for $I=0$ and $\sum_{q=u,d,s}\bar sq\bar qu$ for $I=1/2$. The prime in  Eqs. (\ref{basis_zero},\ref{basis_two}) indicates that two currents may have different flavor structure. The diquark anti-diquark interpolators have flavor structure $[\bar u\bar d] [ud]$ for $I=0$ and $[\bar s\bar d][du]$ for $I=1/2$.} are taken to get $I=0$ and $I=1/2$.

We use $r=3$ interpolators  for $I=2$ and $I=3/2$
 \begin{equation}
\label{basis_two}
{\cal O}_1=PP^\prime~,\quad {\cal O}_2=\sum_{i=1,2,3}V_iV^\prime_i~,\quad {\cal O}_3=\sum_{i=1,2,3}A_iA^\prime_i
\end{equation}
with flavor content $\bar d u\bar d u$ for $I=2$ and $\bar s u\bar d u$ for $I=3/2$. 

We compute the $r\times r$ correlation matrix 
\begin{equation}
\label{c}
C_{ij}(t)=\sum_{\vec x}e^{i\vec p \vec x}\langle 0|{\cal O}_i(\vec x,t){\cal O}_j^\dagger(\vec 0,0)|0\rangle_{\vec p=0}=\sum_n Z_i^{(n)}Z_j^{(n)*}e^{-E^{(n)}t}~
\end{equation}
where $Z_i^{(n)}\equiv \langle 0|{\cal O}_i|n\rangle$. 
 Like in all previous tetraquark simulations, 
we omit the single and double disconnected quark contractions for $I=0,1/2$.  This approximation discards  $\bar qq\bar q q\leftrightarrow \bar qq\leftrightarrow vac$  mixing.  Since we are interested in an intermediate tetraquark state with four valence quarks, there is even a good excuse to use this approximation in these pioneering studies.

The extraction of the energies from the correlation functions (\ref{c}) using a multi-exponential fit is unstable. Instead, we solve the generalized eigenvalue problem \cite{var}
\begin{equation}
C(t)\vec u^{(n)}(t)=\lambda^{(n)}(t,t_0) C(t_0)\vec u^{(n)}(t) ~,
\end{equation} 
which gives us the energies $E^{(n)}$ via $\lambda^{(n)}(t)$ and the ratios of $Z_i^{(n)}$  via $\vec u^{(n)}(t)$ as follows:
\begin{equation}
\label{lam}
\lambda^{(n)}(t)\simeq  e^{-E^{(n)}(t-t_0)}~,\qquad  \frac{|\sum_k C_{ik}(t) u_k^{n}(t)|}{|\sum_{k^\prime} C_{jk^\prime}(t) u_{k^\prime}^{n}(t)|}\simeq \bigg\vert\frac{Z_i^{(n)}}{Z_j^{(n)}}\bigg\vert~.
\end{equation}
These relations are formally valid at large $t$, infinite lattice temporal extent $T$ and $t_0<t\leq 2t_0$ \cite{var}. 

\section{Results for $I=0$ and $I=2$}

\begin{figure}[htb!]
\begin{center}
\includegraphics[height=5.2cm,clip]{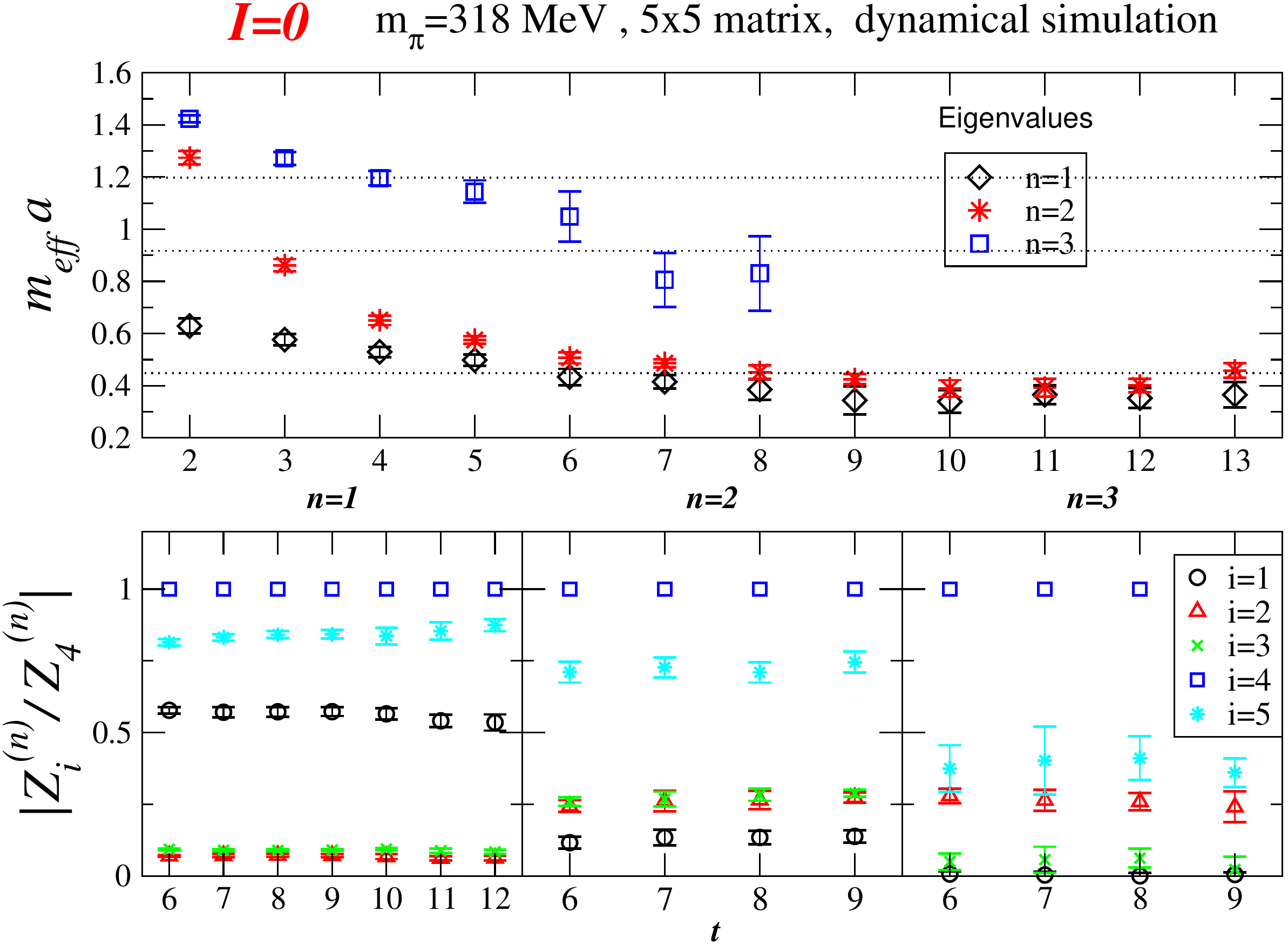} 
\includegraphics[height=5.2cm,clip]{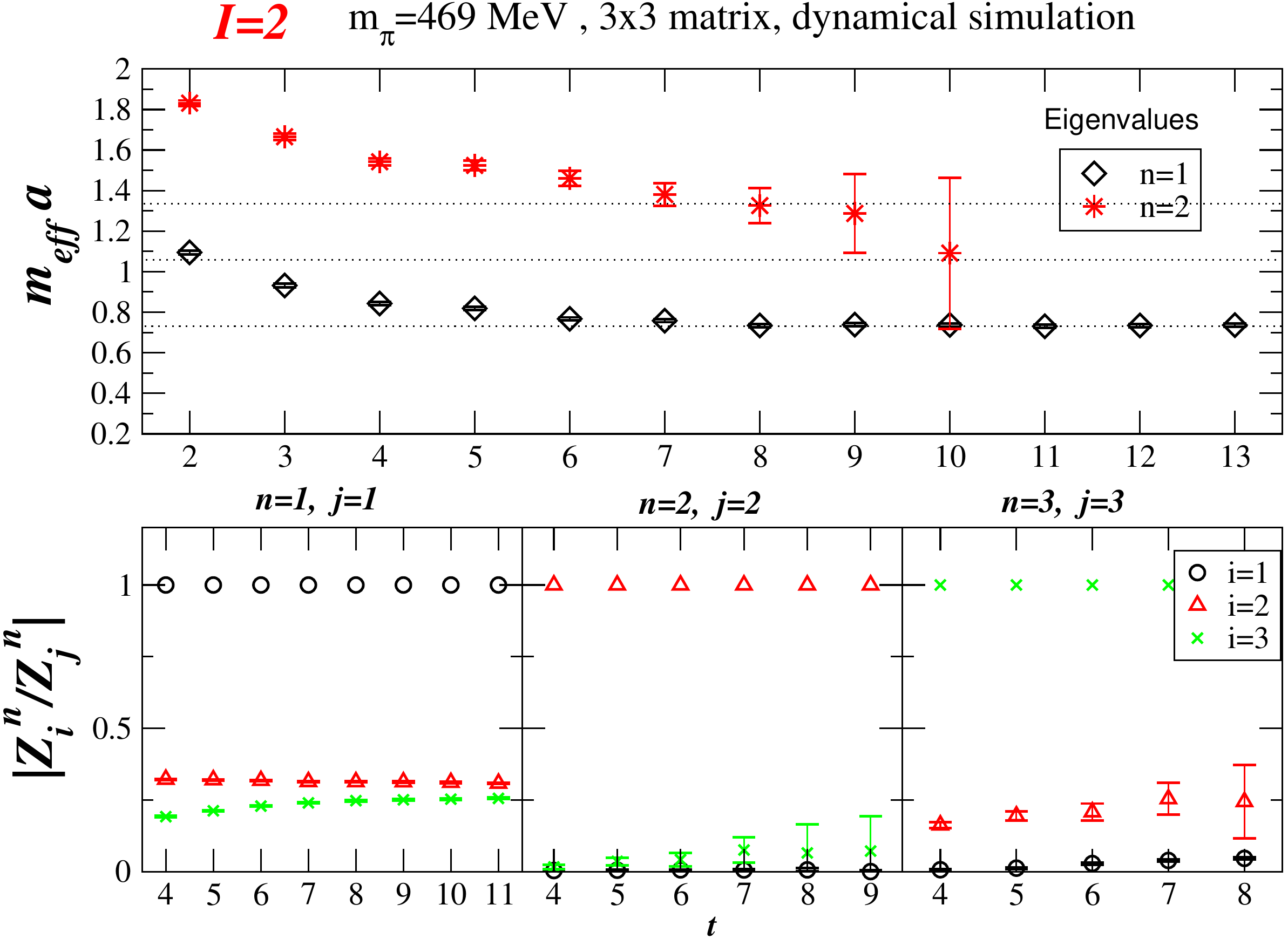}
\end{center}
\caption{ \small  Typical effective masses of the eigenvalues $\lambda^{(n)}(t)$ for $I=0,~2$ and $n=1,~2,~3$. Corresponding ratios  $|Z_i^{(n)}/Z_j^{(n)}|$ at given $n$  are also shown ($j$ is the largest component).  Results for specific  quark masses and $t_0=1$ in the dynamical simulation are shown. The  lines present the energies of non-interacting $\pi(k)\pi(-k)$ with $k=N\tfrac{2\pi}{L}$ and $N=0,1,\sqrt{2}$. }\label{zero_two_eff}
\end{figure}

The typical effective masses and the ratios $|Z_i^{(n)}/Z_j^{(n)}|$ for $I=0,~2$  are given in Fig. \ref{zero_two_eff}.  The lines display the three lowest energies of $\pi(k)\pi(-k)$ in the non-interacting case. 

\begin{figure}[htb!]
\begin{center}
\includegraphics[height=5.1cm,clip]{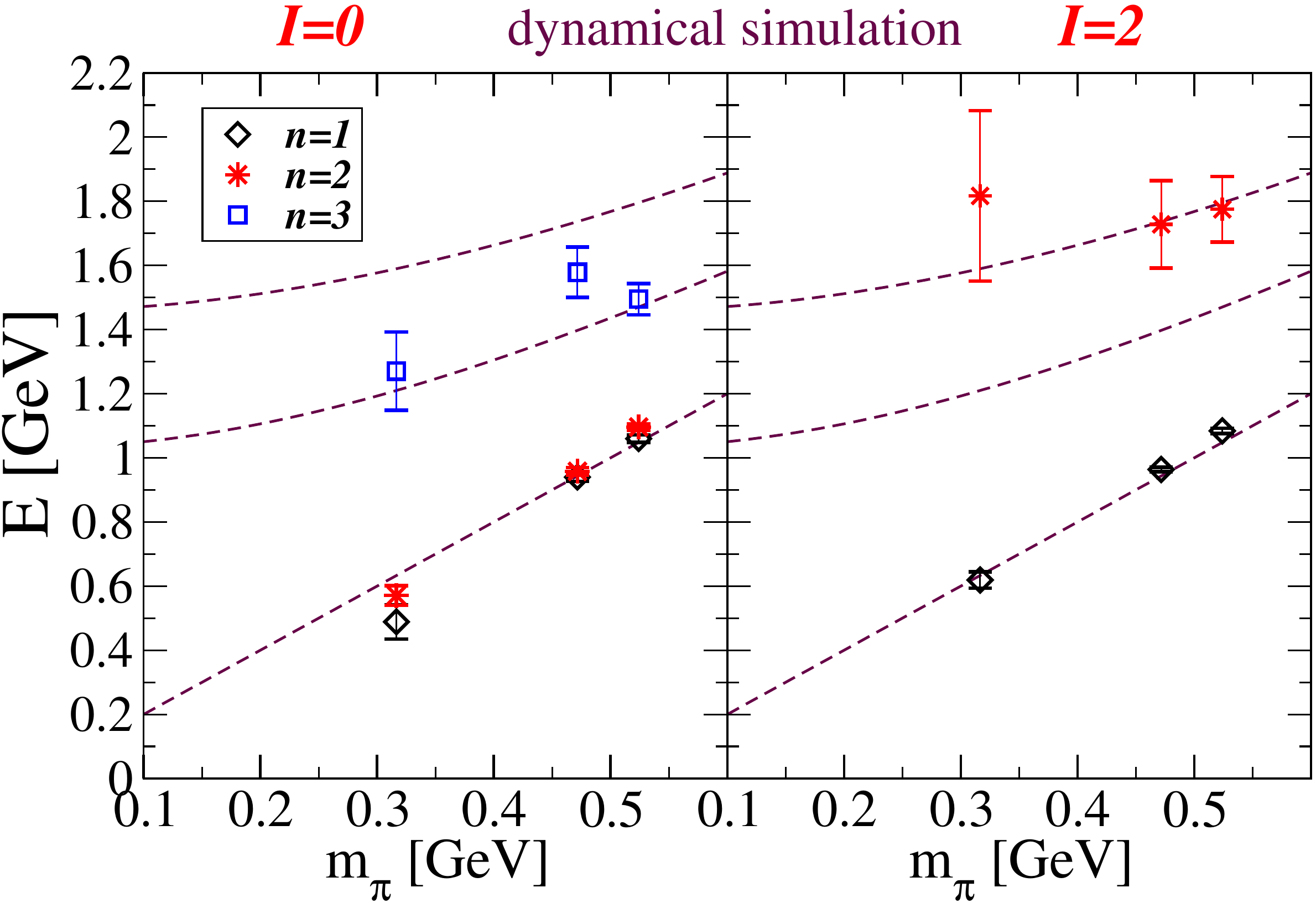}
\includegraphics[height=5.1cm,clip]{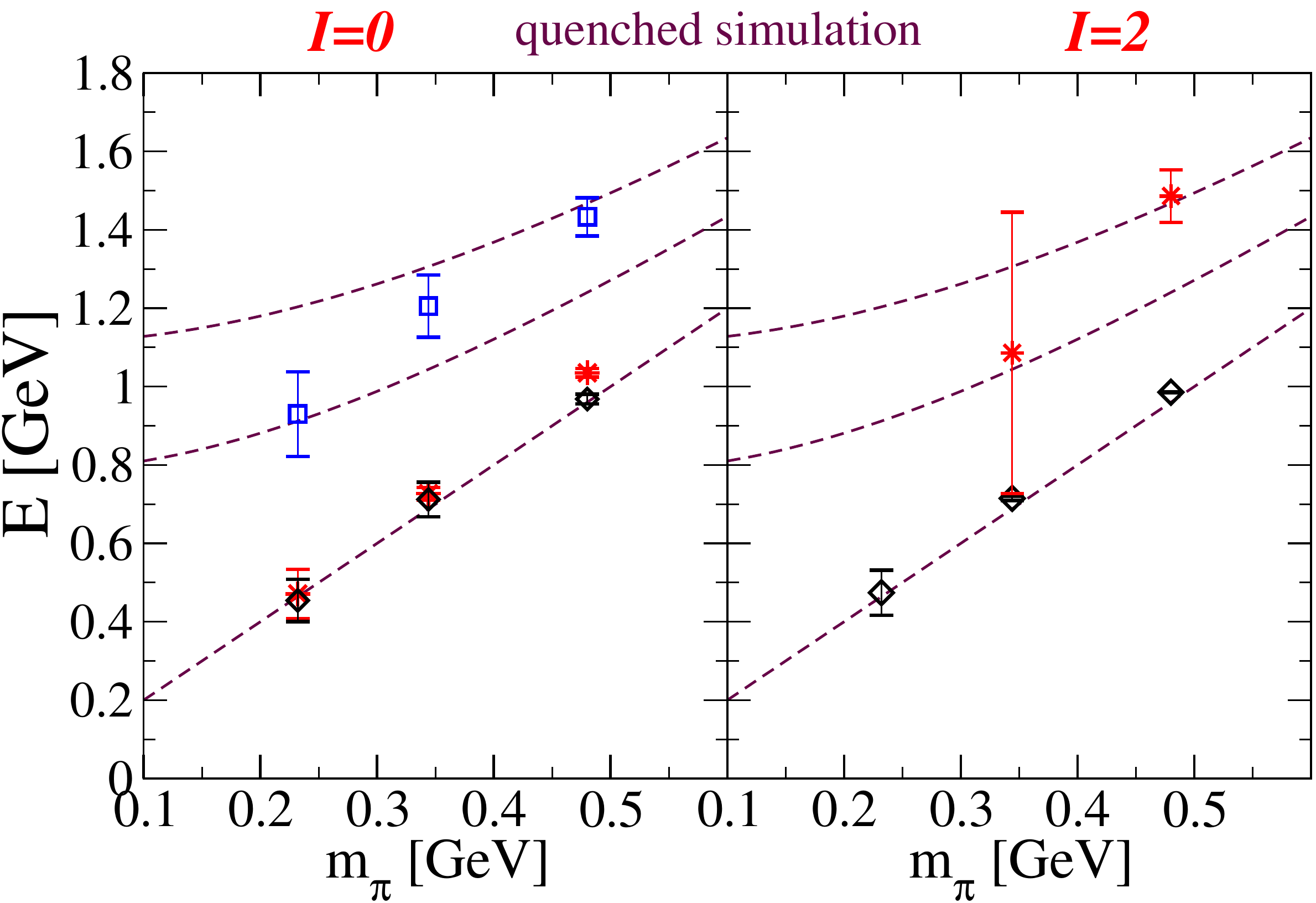}
\end{center}
\caption{ \small The resulting spectrum $E^{(n)}$ for $I=0$ and $I=2$ in the dynamical (left) and the quenched (right) simulations. Note that there are two states (black and red) close to each other in $I=0$ case. The lines present the energies of non-interacting $\pi(k)\pi(-k)$ with $k=N\tfrac{2\pi}{L}$ and $N=0,1,\sqrt{2}$. }\label{zero_two_spect}
\end{figure}

In $I=0$ case, we find one state with energy close to $\pi(0)\pi(0)$, another state with energy close to $\pi(\tfrac{2\pi}{L})\pi(-\tfrac{2\pi}{L})$ and we also find an additional light state (close to the lowest state). This applies for all quark masses and for dynamical as well as the quenched simulation, as shown in the resulting spectrum on Fig. \ref{zero_two_spect}. Both figures display results for  $t_0=1$ and for the case when the full $5\times 5$ matrix (\ref{basis_zero}) was diagonalized. 
We also used  $t_0\in [2,4]$ and diagonalization of all possible $4\times 4$ and $3\times 3$ sub-matrices and we find that extracted $E^{(n)}$ and  $|Z_i^{(1)}/Z_j^{(1)}|$ are almost independent on these choices. The ratios $|Z_i^{(n>1)}/Z_j^{(n>1)}|$ for excited states depend only mildly on these choices.

In $I=2$ case, we find one state with energy close to $\pi(0)\pi(0)$, another state with energy close to $\pi(\tfrac{2\pi}{L})\pi(-\tfrac{2\pi}{L})$ and no additional light state (see Figs. \ref{zero_two_eff} and \ref{zero_two_spect}). Again, this applies for all quark masses, both simulations and for the range of $t_0\in [1,4]$. In this case we use only a $3\times 3$ matrix (\ref{basis_two}), which is probably not large enough to capture the energy of the $\pi(\tfrac{2\pi}{L})\pi(-\tfrac{2\pi}{L})$ exactly (it naturally comes out to high). We point out that our intention was not to capture energy of $\pi(\tfrac{2\pi}{L})\pi(-\tfrac{2\pi}{L})$ correctly, but to verify that there is no light state in addition to $\pi(0)\pi(0)$ in the $I=2$ channel.

\vspace{0.2 cm}

The energies $E^{(n)}$ in Figs. \ref{zero_two_spect} and  \ref{half_threehalf_spect} were fitted from the eigenvalues $\lambda^{(n)}(t)$ using \footnote{The three-parameter fit of the second (first) excited state for $I=0$ ($I=2$) was unstable, so we fixed $A^{(n)}=0$ in this case.}  
\begin{equation}
\label{effect}
\lambda^{(n)}(t)=w^{(n)}\bigl[e^{-E^{(n)}t}+e^{-E^{(n)}(T-t)}\bigr]+A^{(n)}\bigl[e^{-E_{P_1}t}e^{-E_{P_2}(T-t)}+e^{-E_{P_2}t}e^{-E_{P_1}(T-t)}\bigr]~,
\end{equation}
which takes into account the effect of finite $T$ and periodic boundary conditions for a scattering state $P_1P_2=\pi\pi,~K\pi$ (see Appendix A of \cite{prelovsek}). The time-dependence of the ground state $\lambda^{(1)}(t)$  is in   good agreement with the expectation (\ref{effect}) and the fit typically renders $A^{(n)}\simeq w^{(n)}$. The effective masses, plotted in Figs. \ref{zero_two_eff} and \ref{half_threehalf_eff}, are also defined by taking into account the time-dependence (\ref{effect}) \cite{prelovsek}.

\section{Results for $I=1/2$ and $I=3/2$}

\begin{figure}[htb!]
\begin{center}
\includegraphics[height=5.2cm,clip]{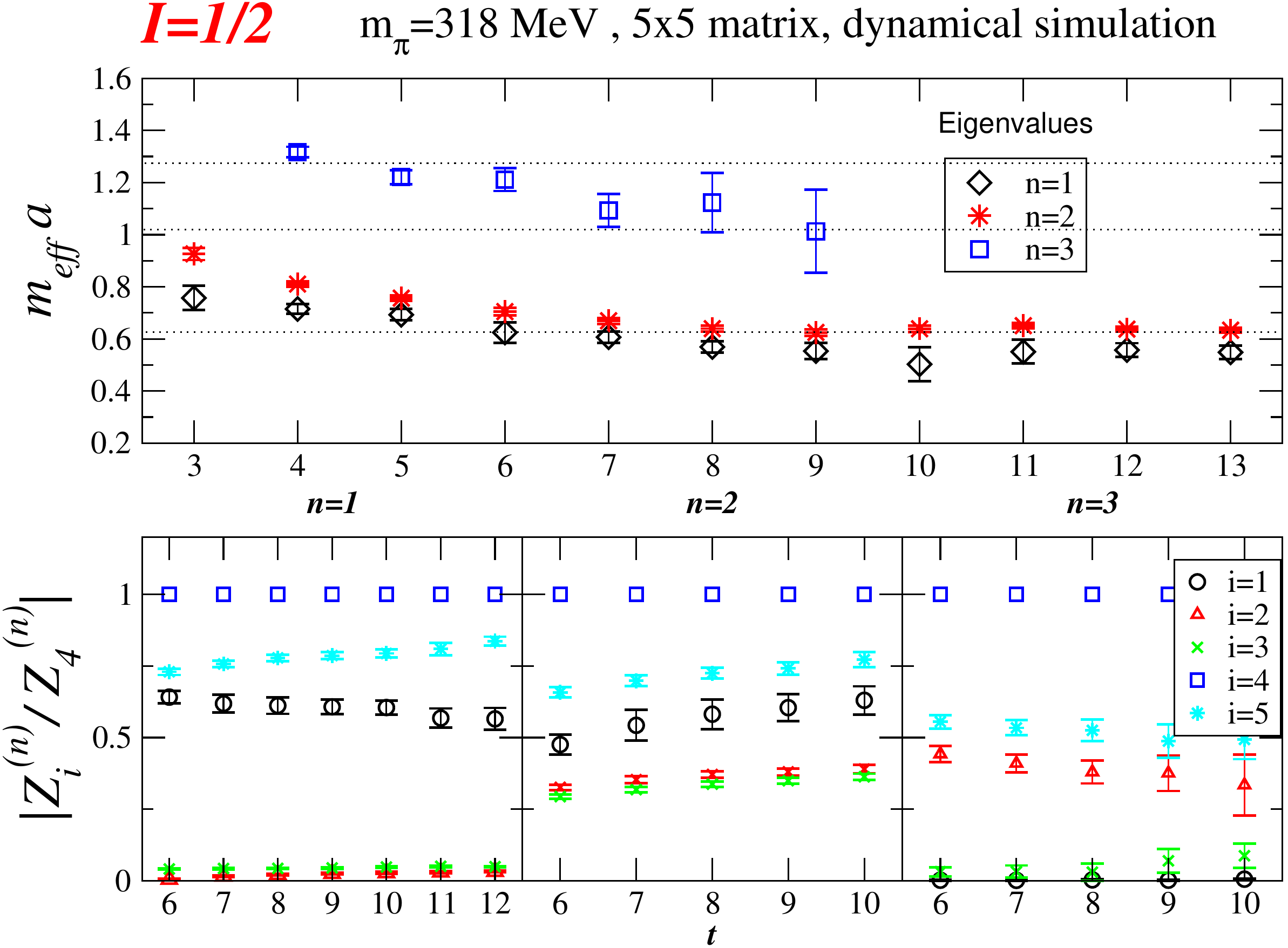} 
\includegraphics[height=5.2cm,clip]{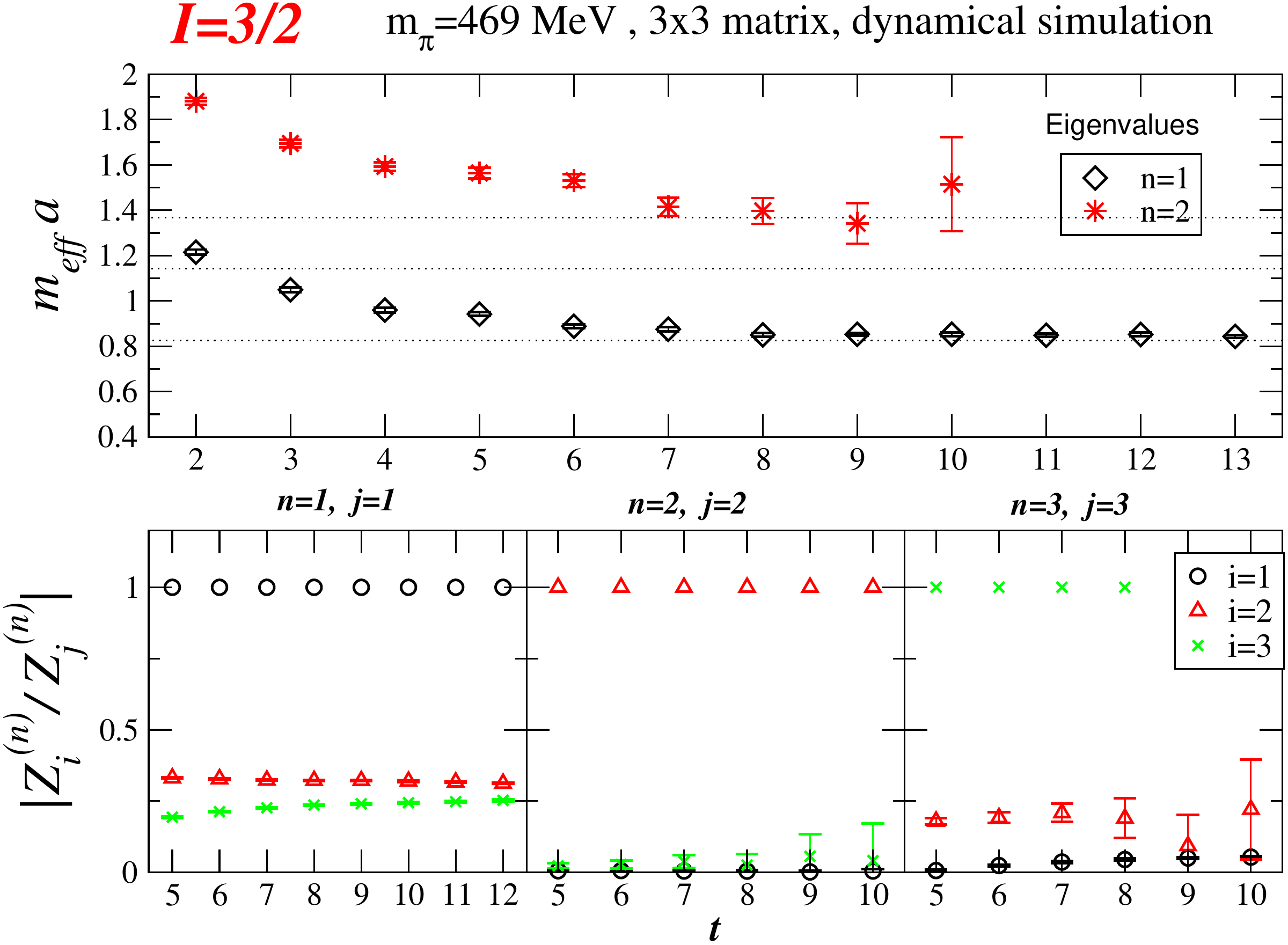}
\end{center}
\caption{ \small Typical effective masses of the eigenvalues $\lambda^{(n)}(t)$ for $I=1/2,~3/2$ and $n=1,~2,~3$.  Corresponding ratios  $|Z_i^{(n)}/Z_j^{(n)}|$ at given $n$  are also shown ($j$ is the largest component). Results for specific  quark masses and $t_0=1$ in the dynamical simulation are shown. The lines present the energies of non-interacting $K(k)\pi(-k)$ with $k=N\tfrac{2\pi}{L}$ and $N=0,1,\sqrt{2}$. }\label{half_threehalf_eff}
\end{figure}

\begin{figure}[htb!]
\begin{center}
\includegraphics[height=5.1cm,clip]{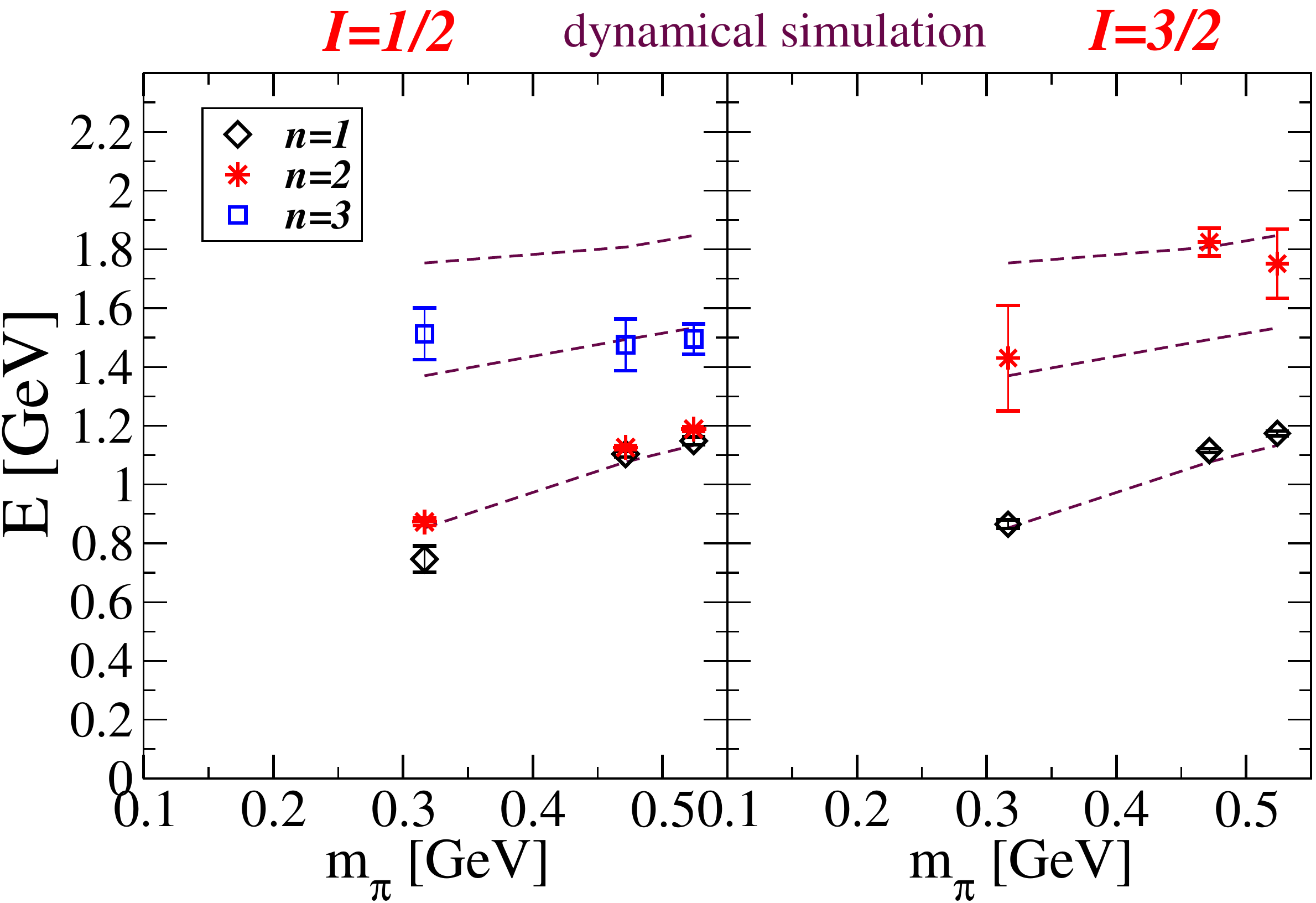}
\includegraphics[height=5.1cm,clip]{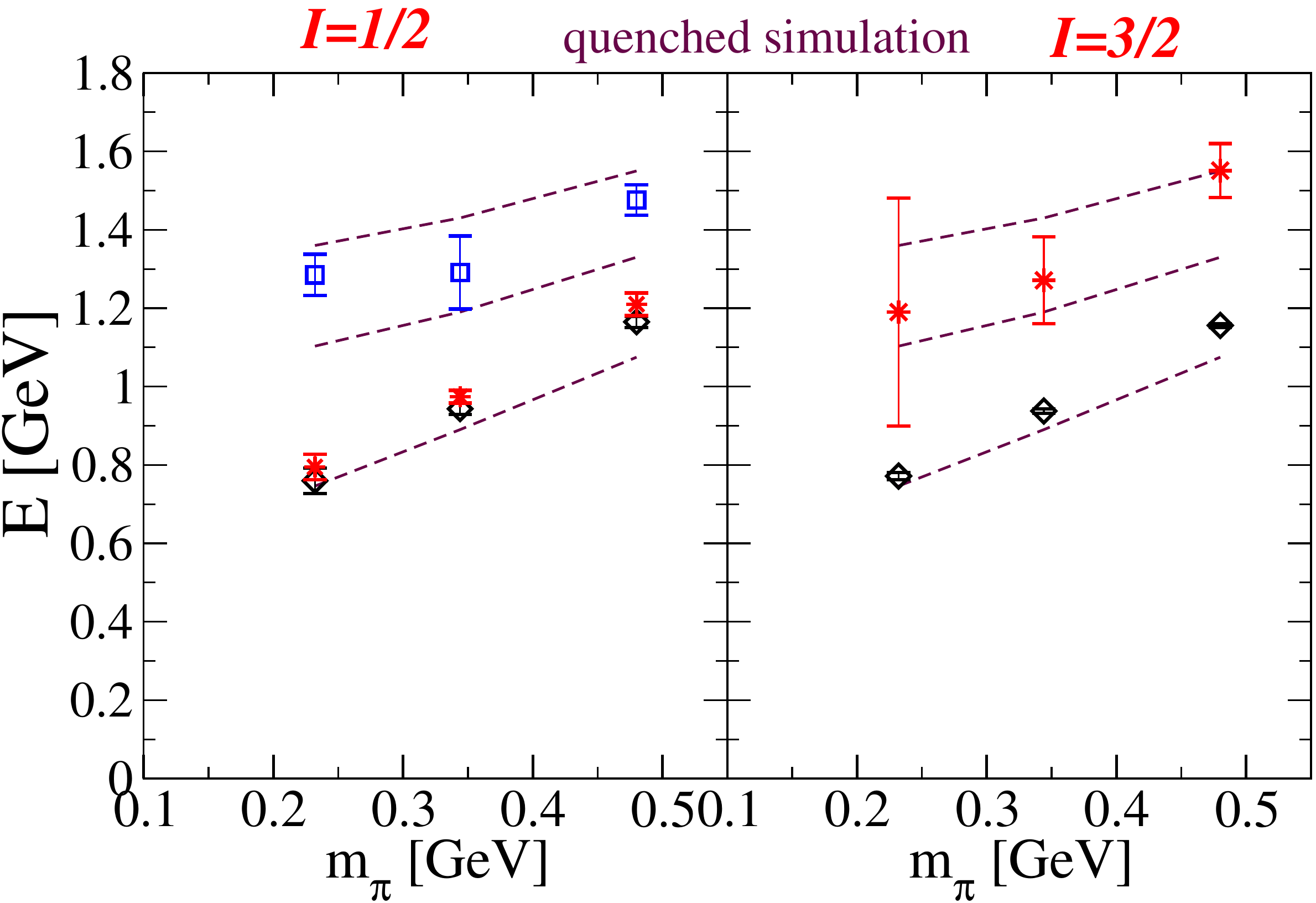}
\end{center}
\caption{ \small The resulting spectrum $E^{(n)}$ for $I=1/2,~3/2$ in the dynamical and the quenched simulation. Note that there are two states (black and red) close to each other in $I=1/2$ case. The lines present the energies of non-interacting $K(k)\pi(-k)$ with $k=N\tfrac{2\pi}{L}$ and $N=0,1,\sqrt{2}$.  }\label{half_threehalf_spect}
\end{figure}

The effective masses and resulting spectrum for $I=1/2,~3/2$ are shown 
in Figs. \ref{half_threehalf_eff} and \ref{half_threehalf_spect}. The conclusions regarding $I=1/2$ is completely analogous to the $I=0$ case above: there is a light state in addition to $K(0)\pi(0)$ and $K(\tfrac{2\pi}{L})\pi(-\tfrac{2\pi}{L})$. Results for the exotic $I=3/2$ channel are analogous to results for  $I=2$: there is no light state in addition to $K(0)\pi(0)$ and $K(\tfrac{2\pi}{L})\pi(-\tfrac{2\pi}{L})$.  

We verified again that these conclusions are independent on the choice of $4\times 4$ or $3\times 3$ sub-matrices of the full $5\times 5$ matrix (for $I=1/2$ case) and they are independent on the choice of $t_0\in [1,4]$.

\section{Conclusions and outlook}

We determined the energy spectrum of states with $J^{PC}=0^{++}$ and $I=0,~1/2,~3/2,~2$ using the generalized eigenvalue method with a number of tetraquark interpolators at the source and the sink. Our simulation is done at several values of $m_\pi$ and  at a single lattice size $L$. Our main question is whether we find any  light states in addition to the towers of $\pi(k)\pi(k)$ or $K(k)\pi(k)$ ($k=0,~\tfrac{2\pi}{L},~..$) scattering states. 

We do find additional light states in the $I=0$ and $I=1/2$ channels. We conclude that these additional states may be possibly related to the observed resonances $\sigma$ and $\kappa$. Since we used only the tetraquark sources and discarded the annihilation diagrams, our simulation also gives  an indication that $\sigma$ and $\kappa$ may have a strong tetraquark component. 

We point out that results from our simulation would have to be confirmed by an independent lattice simulation, before making firm conclusions. At this point, we can not completely exclude the possibility that the additional light states observed in $I=0,~1/2$ channels are some kind of unknown artifact from the generalized eigenvalue method, omission of annihilation diagrams, etc. However, our results behave well in a number of extensive checks we made to exclude this possibility.

The ultimate method to study $\sigma$ and $\kappa$ on the lattice would involve the study of the energy spectrum as a function of the lattice size $L$. The resonance mass and width could then be determined using the L\"{u}scher's finite volume method \cite{luscher}. The importance of the tetraquark component could be determined by comparing $\langle 0|{\cal O}_4|n\rangle$ and   $\langle 0|\bar q q|n\rangle$.  

\vspace{1cm}
    
{\bf Acknowledgments}

\vspace{0.2cm}
 
We would like to thank  C. Gattringer, C. Morningstar, R. Edwards, J. Juge, M. Savage, W. Detmold and M. Komelj for valuable discussions. The configurations with dynamical quarks have been produced by the
BGR-collaboration  on the SGI Altix 4700 of the Leibniz-Rechenzentrum Munich. This work is supported by the Slovenian Research Agency, by the European RTN network FLAVIAnet (contract number MRTN-CT-035482), by the Slovenian-Austrian bilateral project (contract number  BI-AT/09-10-012), by the USA DOE Grant DE-FG05-84ER40154 and by the Austrian grant FWF DK W1203-N08.

\end{document}